\documentstyle[prb,twocolumn,aps,psfig,amssymb,floats]{revtex}

\begin{document}
\draft

\twocolumn[\hsize\textwidth\columnwidth\hsize\csname@twocolumnfalse\endcsname

%%%%%%%%%%%%%%%%%%%%%%%%%%%%%%%%%%%%%%%%%%%%%%%%%%%%%%%%%%%%%%%%%
%%%%%%   TITLE   %%%%%%%%%%%%%%%%%%%%%%%%%%%%%%%%%%%%%%%%%%%%%%%%
%%%%%%%%%%%%%%%%%%%%%%%%%%%%%%%%%%%%%%%%%%%%%%%%%%%%%%%%%%%%%%%%%

\title{The periodic Anderson model with correlated conduction
electrons}
\author{Tom Schork and Stefan Blawid}
\address{Max-Planck-Institut f\"ur Physik komplexer Systeme,\\
Bayreuther Str.\ 40 Haus 16, D-01187 Dresden, Germany}
\date{\today}

\maketitle

%%%%%%%%%%%%%%%%%%%%%%%%%%%%%%%%%%%%%%%%%%%%%%%%%%%%%%%%%%%%%%%%%
%%%%%%   ABSTRACT   %%%%%%%%%%%%%%%%%%%%%%%%%%%%%%%%%%%%%%%%%%%%%
%%%%%%%%%%%%%%%%%%%%%%%%%%%%%%%%%%%%%%%%%%%%%%%%%%%%%%%%%%%%%%%%%

\begin{abstract}
We investigate a periodic Anderson model with interacting
conduction electrons which are described by a Hubbard-type
interaction of strength $U_c$. Within dynamical mean-field theory
the total Hamiltonian is mapped onto an impurity model, which is
solved by an extended non-crossing approximation. We consider the
particle-hole symmetric case at half-filling. Similar to the case
$U_c=0$, the low-energy behavior of the conduction electrons at
high temperatures is essentially unaffected by the $f$ electrons
and for small $U_c$ a quasiparticle peak corresponding to the
Hubbard model evolves first. These quasiparticles screen the $f$
moments when the temperature is reduced further, and the system
turns into an insulator with a tiny gap and flat bands. The
formation of the quasiparticle peak is impeded by increasing
either $U_c$ or the $c$-$f$ hybridization. Nevertheless almost
dispersionless bands emerge at low temperature with an increased
gap, even in the case of initially insulating host electrons. The
size of the gap in the one-particle spectral density at low
temperatures provides an estimate for the low-energy scale and
increases as $U_c$ increases.
\end{abstract}

\pacs{71.10.-w,71.27.+a,75.20.Hr,71.10.Fd}

\vskip2pc]

%%%%%%%%%%%%%%%%%%%%%%%%%%%%%%%%%%%%%%%%%%%%%%%%%%%%%%%%%%%%%%%%%
%%%%%%   INTRODUCTION   %%%%%%%%%%%%%%%%%%%%%%%%%%%%%%%%%%%%%%%%%
%%%%%%%%%%%%%%%%%%%%%%%%%%%%%%%%%%%%%%%%%%%%%%%%%%%%%%%%%%%%%%%%%

\section{Introduction}

The usual explanation for the formation of heavy fermions in
compounds with rare-earth or actinide elements is based on the
Kondo effect.\cite{Fulde88,HewsonBook} Thereby, the characteristic
low-energy scale arises from the spin-screening of the local
moments by a non-interacting electron gas. The periodic Anderson
model is considered as the most promising candidate to at least
qualitatively describe the rich physics of these materials. This
standard scenario, however, fails to explain the heavy-fermion
behavior found in the electron-doped cuprate $\rm
Nd_{2-x}Ce_xCuO_4$ discovered a few years ago.~\cite{Brugger93} In
particular, the estimated low-energy scale is orders of magnitude
too small.\cite{Fulde93} Since undoped $\rm Nd_2CuO_4$ is an
antiferromagnetic charge-transfer
insulator\cite{Skanthakumar89,Oseroff90} despite of one hole per
unit cell, it has been suggested\cite{Fulde93} that this
discrepancy is due to the strong interactions among the electrons
introduced by doping.

As regards the influence of correlated conduction electrons on the
Kondo effect, up to now attention has been focused on the case of
a magnetic impurity embedded in a correlated host which is either
described by a Luttinger liquid in one
dimension\cite{Furusaki94,Li95,Froejdh96} or by some kind of
Hubbard model in higher
dimensions.\cite{Schork94,Khaliullin95,Igarashi95,Igarashi95b,%
Schork96,Wang96b} In all of these cases one finds a strong
dependence of the low-energy scale on the interaction strength of
the conduction electrons and its increase with increasing
interaction strength. For a lattice of moments hybridizing with
correlated electrons only few results
exist.\cite{Shibata96,Itai96}

As a first step towards understanding the effect of conduction
electron interactions on the formation of heavy fermions, we
consider a lattice of $f$-electrons that hybridize with conduction
electrons which themselves are correlated. These correlations will
be described by a Hubbard-type interaction. The resulting model
combines a periodic Anderson model with a Hubbard model. A limit
in which one may obtain sensible results for this locally highly
correlated model is the limit of large spatial
dimensions.\cite{Metzner89,MuellerHartmann89,Georges96} In this
limit the dynamics becomes essentially
local.\cite{MuellerHartmann89} Hence for any correlated model, a
single (correlated) site may be chosen and embedded in an
effective medium which has to be determined self-consistently
(``dynamical mean-field theory''): The model reduces to an
Anderson impurity model.\cite{Ohkawa91,Jarrell92,Georges92b} In
fact, in has been shown that besides the Hubbard
model\cite{Ohkawa91,Ohkawa92,Jarrell92,Georges92} the periodic
Anderson model with uncorrelated conduction electrons is amenable
to this limit.\cite{Ohkawa92c,Jarrell93b,Jarrell95,Saso96}

In the next section we introduce the model and derive the
corresponding impurity model. The impurity model is solved
numerically by an extended non-crossing
approximation\cite{Keiter70,Bickers87,Pruschke89,Lombardo96}
which is derived in Sec.~\ref{sec:model} as well. In
Sec.~\ref{sec:results} we present results for the particle-hole
symmetric case at half filling. Assuming a paramagnetic ground
state we study the influence of weak correlations on the
one-particle spectral density and discuss how the heavy bands
emerge. At low temperatures a gap forms which in the free case is
related to the Kondo temperature\cite{Jarrell95} and we discuss
how the size of the gap depends on the strength of the
correlations of the conduction electrons. We finally conclude in
Sec.~\ref{sec:conclusions}.

%%%%%%%%%%%%%%%%%%%%%%%%%%%%%%%%%%%%%%%%%%%%%%%%%%%%%%%%%%%%%%%%%
%%%%%%   MODEL AND METHOD   %%%%%%%%%%%%%%%%%%%%%%%%%%%%%%%%%%%%%
%%%%%%%%%%%%%%%%%%%%%%%%%%%%%%%%%%%%%%%%%%%%%%%%%%%%%%%%%%%%%%%%%

\section{Model and Method}
\label{sec:model}

%%%%%%%%%%%%%%%%%%%%%%%%%%%%%%%%%%%%%%%%%%%%%%%%%%%%%%%%%%%%%%%%%
%%%%%%   LATTICE   %%%%%%%%%%%%%%%%%%%%%%%%%%%%%%%%%%%%%%%%%%%%%%
%%%%%%%%%%%%%%%%%%%%%%%%%%%%%%%%%%%%%%%%%%%%%%%%%%%%%%%%%%%%%%%%%

\subsection{The lattice Hamiltonian}

In the following we consider the simplest version of the periodic
Anderson model and allow for interacting conduction electrons,
\begin{eqnarray}
H & = & H_c + H_f + H_{cf} 
\label{hamil} \\
H_c & = & 
\sum_{k\sigma} (\epsilon_k+\epsilon_c-\mu) 
c^\dagger_{k,\sigma} c^{\phantom{\dagger}}_{k,\sigma} 
+ U_c \sum_i n^c_{i\uparrow} n^c_{i\downarrow}
\label{hamil_c} \\
H_f & = & (\epsilon_f-\mu) \sum_{i\sigma} f^\dagger_{i,\sigma}
f^{\phantom{\dagger}}_{i,\sigma} + U_f \sum_i n^f_{i\uparrow}
n^f_{i\downarrow}
\label{hamil_f} \\
H_{cf} & = & V \sum_{i\sigma} \left( 
f^\dagger_{i,\sigma} c^{\phantom{\dagger}}_{i,\sigma}
+ c^\dagger_{i,\sigma} f^{\phantom{\dagger}}_{i,\sigma}  \right)
~. 
\label{hamil_cf}
\end{eqnarray}
Here, $f_{i\sigma}^{(\dagger)}$ destroys (creates) an electron in
the localized $f$ orbital at site $i$ with spin $\sigma$, and
$U_f$ is the Hubbard interaction of the localized $f$ states
($\epsilon_f<0$). The $c$ operators refer to the conduction
electrons which are described by a Hubbard model with an
interaction $U_c$ being typically smaller than $U_f$. $\mu$
denotes the chemical potential and $V$ measures the mixing
between the $c$ and $f$ subsystems. In the following we will
refer to the Hamiltonian~(\ref{hamil}) as ``(periodic)
Anderson-Hubbard model.''

In what follows we will concentrate on the one-particle spectra
in a paramagnetic phase. Information on the underlying lattice
will enter the dynamical mean-field equations only via the
density of states of the conduction electrons.\cite{Georges96}
For simplicity, we therefore consider a semicircular density of
conduction electrons states with width $2D$ in
Eq.~(\ref{hamil_c})
\begin{equation}
\rho(z) = \frac{2}{\pi D} \sqrt{1-\left(\frac{z}{D}\right)^2}
\label{DOS}
\end{equation}
which arises from hopping on a Bethe-lattice of coordination
number $Z$ with matrix element $t=D/(2\sqrt Z)$ in the limit
$Z\to\infty$. We will use $D=1$ as unit of energy throughout this
paper.

Given the non-interacting Green's functions, $G_0(k,z)$, 
\begin{equation}
G_0^{-1}(k,z) = 
\left( 
\begin{array}{c c}
z-(\epsilon_f-\mu) & -V \\
-V & z-(\epsilon_k-\mu)
\end{array}
\right)
\end{equation}
and the full Green's function $G(k,z)$, the self-energies are
defined by Dyson's equation
\begin{eqnarray}
G^{-1}(k,z) & = & G_0^{-1}(k,z) - \Sigma(k,z)
\label{greens}\\ 
\Sigma(k,z) & = & 
\left( 
\begin{array}{c c}
\Sigma_f(k,z) & \Sigma_{fc}(k,z) \\
\Sigma_{cf}(k,z) & \Sigma_c(k,z)\\
\end{array}
\right)
~.
\label{self}
\end{eqnarray}

%%%%%%%%%%%%%%%%%%%%%%%%%%%%%%%%%%%%%%%%%%%%%%%%%%%%%%%%%%%%%%%%%
%%%%%%   IMPURITY   %%%%%%%%%%%%%%%%%%%%%%%%%%%%%%%%%%%%%%%%%%%%%
%%%%%%%%%%%%%%%%%%%%%%%%%%%%%%%%%%%%%%%%%%%%%%%%%%%%%%%%%%%%%%%%%

\subsection{The impurity model}

The dynamical mean-field theory assumes that the self-energy is a
local quantity which is correct in the limit of infinite
dimensions.\cite{Metzner89,MuellerHartmann89,Ohkawa92,Georges96}
The lattice model can then be mapped onto an impurity
model\cite{Georges96} which is seen as follows: The self
energy~(\ref{self}) of the lattice model~(\ref{hamil}) is given
as the derivative of a functional $\Phi[G]$ of the full Green's
functions:\cite{AbrikosovBook}
\begin{equation}
\Sigma_{\mu\nu}(i,j;z) = 
\frac{\delta\Phi}{\delta G_{\nu\mu}(j,i;z)}~.
\end{equation}
Here, $i$ and $j$ are real space coordinates and $\mu,\nu$
correspond to $c$ and $f$. If the self-energy of the lattice
model is local, $\Phi$ depends on the local Green's functions,
$G(i,i)$, only. Thus, $\Phi$ can be generated from an impurity
model. Solving the impurity model we know $\Phi$ and, hence, the
self-energy as functional of the impurity Green's function ${\cal
G}$: $\Sigma[{\cal G}] = \delta\Phi / \delta {\cal G}$. We now
identify ${\cal G}$ with the local Green's function of the
lattice,
\begin{equation}
G_{\rm loc}(z) = \frac{1}{N} \sum_k G(k,z) 
= \int d\epsilon~ \rho(\epsilon) G(\epsilon,z) ~.
\label{loc_gf}
\end{equation}
Note that the $k$-dependence enters only via $\epsilon_k$ into
$G(k,z)$, thus $G_{\rm loc}(z)$ can be expressed by an energy
integration. From $G_{\rm loc} = {\cal G}$, we find the actual
value of $\Sigma$ from the functional equation
\begin{eqnarray}
G_{\rm loc}(z) & = & \int d\epsilon~ \rho(\epsilon) G(\epsilon,z)
\nonumber \\ & = & \int d\epsilon~\rho(\epsilon)
\left(G_0^{-1}(\epsilon,z) - \Sigma[G_{\rm loc}]\right)^{-1}~.
\label{cond}
\end{eqnarray}
Technically, Eq.~(\ref{cond}) determines the free Green's
function of the impurity model which is of course not fixed by
$\Phi$.

The Hamiltonian of the impurity model that generates $\Phi$ is
not unique. Since $\Phi$ is the same for both impurity and
lattice model, they have the same diagrammatic expansion. We
therefore embed a single unit cell (a ``$c$-$f$ molecule'') as
impurity ($H_{\rm loc}$) in an effective medium ($H_{\rm med}$)
which will be determined self-consistently:
\begin{eqnarray}
H_{\rm imp} & = & H_{\rm loc} + H_{\rm med} 
\label{hamil_imp}
\\
H_{\rm loc} & = & 
\tilde\epsilon_c \sum_\sigma 
c^\dagger_{\sigma} c^{\phantom{\dagger}}_{\sigma}
+ U_c n^c_\uparrow n^c_\downarrow 
+ V \sum_\sigma 
\left( c^\dagger_\sigma f^{\phantom{\dagger}}_\sigma
+ {\rm H.c.} \right)
\nonumber \\
&& + \tilde\epsilon_f 
\sum_\sigma f^\dagger_{\sigma} f^{\phantom{\dagger}}_{\sigma}
+ U_f n^f_\uparrow n^f_\downarrow 
\\
H_{\rm med} & = & 
\sum_{k\sigma} E_k 
\alpha^\dagger_{k\sigma}\alpha^{\phantom{\dagger}}_{k\sigma}
+ \sum_{k\sigma} \left( W_k c^\dagger_\sigma
\alpha^{\phantom{\dagger}}_{k\sigma} + {\rm H.c.} \right) ~.
\end{eqnarray}
(Formally one can consider the action of the lattice model and
integrate out the non-local part to arrive at an impurity action
which is afterwards modelled by an Hamiltonian.\cite{Georges96})
Note that this choice for the impurity model differs
qualitatively from the usual one for the periodic Anderson model
at $U_c=0$.\cite{Georges92b,Ohkawa92c,Jarrell93b} In the case of
free conduction electrons, only a single self energy exists (for
the $f$-electrons) and, hence, only the $f$-orbital is coupled to
an effective medium. When the conduction electrons are correlated
they have a self energy, as well. Therefore, we include a
$c$-orbital in the local part of $H_{\rm imp}$. We need only a
single effective medium, $H_{\rm med}$, although there are two
Hubbard interactions in the original
Hamiltonian~(\ref{hamil}). This is due to the absence of direct
$f$-$f$ hopping: The electrons explore their environment only via
the $c$-orbitals. The two interactions merely show up in the
internal structure of the impurity which consists of two
orbitals, one of which couples to the medium. This situation is
similar to the one encountered in the extended Hubbard model in
Ref.~\onlinecite{Si93}.

We show now that the self-consistency equation (\ref{cond}) can
be fulfilled by our impurity model (\ref{hamil_imp}). The
impurity Green's functions
\begin{equation}
{\cal G}(z) = \left(
\begin{array}{c c}
{\cal G}_f(z) & {\cal G}_{fc}(z) \\
{\cal G}_{cf}(z) & {\cal G}_c(z)
\end{array}
\right)
\end{equation}
are given by
\begin{equation}
\left( 
\begin{array}{c c}
\omega-\tilde\epsilon_f-\Sigma_f(z) & -V-\Sigma_{cf}(z) \\
-V-\Sigma_{fc}(z) & 
\omega -\tilde\epsilon_c -\tilde\Delta(z) -\Sigma_c(z)
\end{array}
\right)
{\cal G}(z) = 1~,
\end{equation}
where
\begin{equation}
\tilde\Delta(z) = \sum_k \frac{|W_k|^2}{z-E_k} ~.
\end{equation}
Equating $\cal{G}$ to the local Green's function $G_{\rm loc}$ of
the lattice model (\ref{loc_gf}), which one obtains analytically
due to the density of states (\ref{DOS}), and using that both,
$\tilde\Delta(z)$ and $G(z) \sim 1/z$ at large $|z|$, we find:
\begin{eqnarray}
\tilde\epsilon_f & = & \epsilon_f - \mu \\
\tilde\epsilon_c & = & \epsilon_c - \mu \\
\tilde\Delta(z) & = & \frac{1}{4} G_{{\rm loc},c}(z) ~.
\end{eqnarray}
The medium is determined by the $c$-Green's function only,
reflecting that there is no direct $f$-$f$ hopping.

The parameters in the impurity model~(\ref{hamil_imp}) is thus
fixed, in particular the medium $\tilde\Delta(z)$ is determined
by the solution of the impurity model ${\cal G}_c(z) = G_{{\rm
loc},c}(z)$.

%%%%%%%%%%%%%%%%%%%%%%%%%%%%%%%%%%%%%%%%%%%%%%%%%%%%%%%%%%%%%%%%%
%%%%%%   SOLVING ...   %%%%%%%%%%%%%%%%%%%%%%%%%%%%%%%%%%%%%%%%%%
%%%%%%%%%%%%%%%%%%%%%%%%%%%%%%%%%%%%%%%%%%%%%%%%%%%%%%%%%%%%%%%%%

\subsection{Solving the impurity model} 

We solve the impurity model (\ref{hamil_imp}) by extending the
non-crossing approximation (NCA)\cite{Keiter70,Bickers87} to the
case of more than two ionic
propagators.\cite{Pruschke89,Lombardo96} This approach has been
applied successfully to the finite-$U$ impurity Anderson
model\cite{Pruschke89} where it has been shown that neglecting
vertex corrections slightly underestimates the Kondo temperature,
and to the Emery model within the dynamical mean-field
theory.\cite{Lombardo96}

Denoting the eigenstates of the local part, $H_{\rm loc}$ by
$|m\rangle$ (with $n_m=0\dots 4$ particles), the impurity
Hamiltonian~(\ref{hamil_imp}) is expressed in terms of Hubbard
operators $X_{mn} = |m\rangle\langle n|$ as
\begin{eqnarray}
H_{\rm loc} & = & \sum_{m=1}^{16} E_m X_{mm} ~. \\
H_{\rm med} & = & 
\sum_{k\sigma} E_k \alpha^\dagger_{k\sigma}\alpha_{k\sigma}
\nonumber\\
&& + \sum_{k\sigma,mn} 
\left( W_k U^c_{mn\sigma} X_{mn} \alpha_{k\sigma}
+  {\rm H.c.} \right)
\end{eqnarray}
with $U^c_{mn\sigma} = \langle m | c^\dagger_\sigma | n \rangle$.

For each state $|m\rangle$ a ionic propagator is introduced
\begin{equation}
R_m(z) = \frac{1}{z-E_m-S_m(z)}
\label{ion_prop}
\end{equation}
with spectral density $\rho_m(x) = -{\rm Im~}
R_m(z+i0^+)/\pi$. We assume that the corresponding self-energies
$S$ are diagonal in the local basis and evaluate them in
self-consistent perturbation theory to second order in the
hybridization $W$ as in the usual NCA:
\begin{eqnarray}
S_m(z) & = & \sum_{n,\sigma} 
(|U^c_{mn\sigma}|^2 + |U^c_{nm\sigma}|^2)  \times
\nonumber \\ 
&& \quad
\int_{-\infty}^{\infty} d\epsilon~
f(\eta_{mn}\epsilon) \Delta(\epsilon) R_n(z+\eta_{mn}\epsilon)
~.
\label{self_nca}
\end{eqnarray}
Here $\eta_{mn} = -1 (+1)$ if the particle number in $|m\rangle$
is higher (lower) than in $|n\rangle$, $f(z) = [\exp(\beta z)+
1]^{-1}$ is the Fermi function, and $\Delta = -{\rm Im~}
\tilde\Delta/\pi$. The $c$- and $f$-Green's functions are given
by
\begin{equation}
G_{c(f)}(z) = \sum_{mn} |U^{c(f)}_{mn,\uparrow}|^2 
\langle\langle X_{mn}; X_{nm} \rangle\rangle_z
\end{equation}
Within the NCA the Green's functions $\langle\langle X_{mn};
X_{nm} \rangle\rangle_z$ are expressed by the ionic
propagators\cite{Bickers87}
\begin{eqnarray}
\lefteqn{\langle\langle X_{mn}; X_{nm} \rangle\rangle_z = }\\
&& \frac{1}{Z} \int dx~ e^{-\beta x} 
[\rho_m(x)R_n(x+z) - \rho_n(x)R_m(x-z)] 
\end{eqnarray}
and 
\begin{equation}
Z = \sum_m \int dx~ e^{-\beta x} \rho_m(x) ~.
\end{equation}

In the symmetric, half-filled case there are only 6 independent
propagators due to particle-hole and spin symmetry. The coupled
integral equations (\ref{ion_prop}) and (\ref{self_nca}) are
solved numerically by introducing defect
propagators\cite{Bickers87b} and making use of the fast Fourier
transformation.\cite{Lombardo96}

%%%%%%%%%%%%%%%%%%%%%%%%%%%%%%%%%%%%%%%%%%%%%%%%%%%%%%%%%%%%%%%%%
%%%%%%   RESULTS   %%%%%%%%%%%%%%%%%%%%%%%%%%%%%%%%%%%%%%%%%%%%%%
%%%%%%%%%%%%%%%%%%%%%%%%%%%%%%%%%%%%%%%%%%%%%%%%%%%%%%%%%%%%%%%%%

\section{Results}
\label{sec:results}

In the following we consider the symmetric model ($\epsilon_c =
-U_c/2$, $\epsilon_f = -U_f/2$) at half-filling
($n_c+n_f=2$). Due to particle-hole symmetry the chemical
potential is 0.

We chose $U_f = 5$ in all our calculations so that the $f$-level
is well outside of the conduction band. Our investigations were
restricted mostly to those values of $U_c$ for which the Hubbard
model for the conduction electrons is metallic (for the
semielliptic density of states (\ref{DOS}) we found $U_{\rm crit}
\sim 1.8$). The reason is that in deriving the self-consistency
equations for the impurity model we assumed a paramagnetic
state. Thus we typically chose $U_c = 0.5 \dots 2.0$ and $V = 0.1
\dots 0.4$. These values lead to small exchange couplings $J\sim
0.01 \dots 0.2$ between two-particle singlet and triplet state of
the $c$-$f$ molecule (see below). We deliberately chose these
small values for $V$ in order to obtain ``Kondo temperatures''
which are small compared to the bare band-width, see
Table~\ref{tab:TKondo}.
\begin{table}
\begin{center}
\begin{tabular}{c|c|c}
$V$ & $T_{\rm K}^{\rm imp}$ & $T_{\rm K}^{\rm lat}$ \\
\hline
0.1 & $ 1.3\times 10^{-44} $ & $ 2.7\times 10^{-23}$ \\
0.2 & $ 2.5\times 10^{-12} $ & $ 5.3\times 10^{-5}$ \\
0.3 & $ 3.1\times 10^{-6} $ & $ 7.3\times 10^{-4}$ \\
0.4 & $ 4.9\times 10^{-4} $ & $ 1.1\times 10^{-2}$
\end{tabular}
\caption{The estimated Kondo temperatures for uncorrelated
conduction electrons. $T_{\rm K}^{\rm imp}$ refers to the
symmetric impurity Anderson model,\protect{\cite{Tsvelick83}}
$T_{\rm K}^{\rm lat}$ includes the lattice enhancement factor of
2 in the exponent for the lattice
case.\protect{\cite{Rice86,Jarrell95}}}
\label{tab:TKondo}
\end{center}
\end{table}

%%%%%%%%%%%%%%%%%%%%%%%%%%%%%%%%%%%%%%%%%%%%%%%%%%%%%%%%%%%%%%%%%
%%%%%%   MOLECULE   %%%%%%%%%%%%%%%%%%%%%%%%%%%%%%%%%%%%%%%%%%%%%
%%%%%%%%%%%%%%%%%%%%%%%%%%%%%%%%%%%%%%%%%%%%%%%%%%%%%%%%%%%%%%%%%

\subsection{Single-particle spectrum of the molecule}
\label{subsec:local}

The following discussions focus on the one-particle spectral
function. It is instructive to investigate this quantity first
for the local problem given by $H_{\rm loc}$ in
Eq.~(\ref{hamil_imp}) in the symmetric case. Due to symmetry, we
consider the photoemission spectrum only. Denoting the
eigenstates and eigenenergies of $H_{\rm loc}$ by $|m\rangle$ and
$E_m$, the one-particle Green's function and spectral function
are given by ($\mu=0$)
\begin{eqnarray}
G_{c}(z) & = & \frac{1}{Z} \sum_{mn} e^{-\beta E_m} \times 
\nonumber \\
&& \left( 
\frac{\langle m|c_\uparrow |n \rangle
\langle n|c^\dagger_\uparrow |m\rangle}
{z-(E_n-E_m)} 
+ \frac{\langle m|c^\dagger_\uparrow |n\rangle
\langle n|c_\uparrow |m\rangle}
{z+(E_n-E_m)} 
\right) \\
A_c(z) & = & -\frac{1}{\pi} {\rm Im~} G_c(z+i0^+) ~.
\end{eqnarray}
At zero temperature only the ground state contributes to the sum
over $m$. It contains two electrons forming a singlet
\begin{eqnarray} 
|\Psi_{\rm S}\rangle & = & \left[ \frac{C_1}{\sqrt 2} 
(c^\dagger_\uparrow  f^\dagger_\downarrow - 
c^\dagger_\downarrow  f^\dagger_\uparrow)
+ \frac{C_2}{\sqrt 2}
(c^\dagger_\uparrow  c^\dagger_\downarrow - 
f^\dagger_\uparrow  f^\dagger_\downarrow) \right] | 0 \rangle 
\\
E_{\rm S} & = & 
-\frac{U_f+U_c}{4} - \sqrt{ \frac{(U_f+U_c)^2}{16} + 4V^2} ~.
\end{eqnarray}
The photoemission spectrum is obtained by removing a particle,
hence the final states are bonding and antibonding combination of
the $c$- and $f$-orbital. To lowest order in $V$ the transition
energies are given by
\begin{eqnarray}
z_c & = & 
-\frac{U_c}{2} - 2\frac{3U_f-5U_c}{(U_c+U_f)(U_f-U_c)} V^2 
\nonumber \\
z_f & = & 
-\frac{U_f}{2} - 2\frac{5U_f-3U_c}{(U_c+U_f)(U_f-U_c)} V^2 ~.
\end{eqnarray}
They correspond to the lower Hubbard bands of the $c$- and
$f$-subsystem which are shifted by the hybridization $V$.

The first excited state is the two-electron triplet state (spin
excitation)
\begin{equation}
|\Psi_{\rm T}\rangle =  \frac{1}{\sqrt 2} 
(c^\dagger_\uparrow  f^\dagger_\downarrow + 
c^\dagger_\downarrow  f^\dagger_\uparrow) | 0 \rangle
\end{equation}
with excitation energy
\begin{eqnarray}
\Delta E = E_{\rm T} - E_{\rm S} 
& = & 
\sqrt{\frac{(U_f+U_c)^2}{16} + 4V^2} - \frac{U_f+U_c}{4} 
\nonumber
\\
& \sim & \frac{8V^2}{U_f+U_c} ~.
\label{exc_energy}
\end{eqnarray}
At the temperatures that we investigate in the following ($T=0.5
\cdots 0.001$) only these two states contribute as initial
states, $|m\rangle$. The resulting photoemission spectrum of the
molecule thus consists of two double peaks at $z_c$, $z_c+\Delta
E$, and $z_f$, $z_f+\Delta E$ and the weight of the peaks shifted
by $\Delta E$ goes to 0 as $T\to 0$.

%%%%%%%%%%%%%%%%%%%%%%%%%%%%%%%%%%%%%%%%%%%%%%%%%%%%%%%%%%%%%%%%%
%%%%%%   1-P SPECTRA   %%%%%%%%%%%%%%%%%%%%%%%%%%%%%%%%%%%%%%%%%%
%%%%%%%%%%%%%%%%%%%%%%%%%%%%%%%%%%%%%%%%%%%%%%%%%%%%%%%%%%%%%%%%%

\subsection{Single-particle spectrum of the Anderson-Hubbard
model} 
\label{subsec:DOS}

We first consider the spectral density of the conduction
electrons
\begin{eqnarray}
A_c(z) & = & -\frac{1}{\pi} {\rm Im~} G_{{\rm loc},c}(z+i0^+) 
\nonumber \\
& = & -\frac{1}{\pi} \sum_k {\rm Im~} G_{c}(k,z+i0^+) 
\end{eqnarray}
which corresponds to photoemission and inverse photoemission
spectra.
\begin{figure}
\centerline{\psfig{file=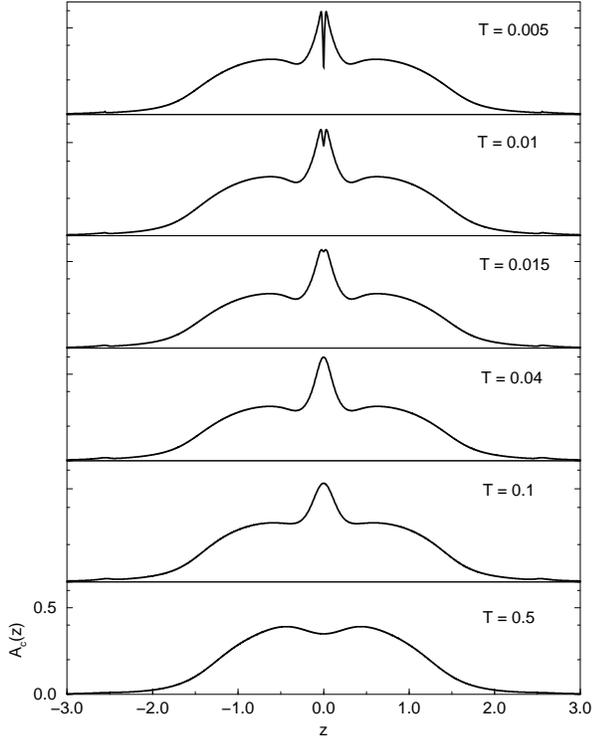,width=8cm}}
\caption{Spectral density $A_c(z)$ for $U_c = 1$, $V = 0.2$, $U_f
= 5$ at different temperatures}
\label{dos_u1v2}
\end{figure}
A typical result for $A_c(z)$ is shown in Fig.~\ref{dos_u1v2} for
different temperatures ($U_c = 1$, $V = 0.2$). At high
temperatures ($T= 0.5$) we obtain two broad maxima located at
$\sim\pm U_c/2$ which correspond to upper and lower Hubbard band
of the $c$-subsystem. When lowering the temperature ($T\lesssim
0.1$), a peak at the chemical potential ($z=0$) arises. This is
the well-known quasiparticle peak belonging to the Hubbard model
of the host electrons, $H_c$.~\cite{Jarrell92,Ohkawa92,Georges92}
In this temperature regime the $f$- and $c$-subsystems are almost
independent. The influence of the $f$-states on the $c$-spectra
is given only by the tiny structure at $z \sim \pm U_f/2 = \pm
2.5$.
\begin{figure}
\centerline{\psfig{file=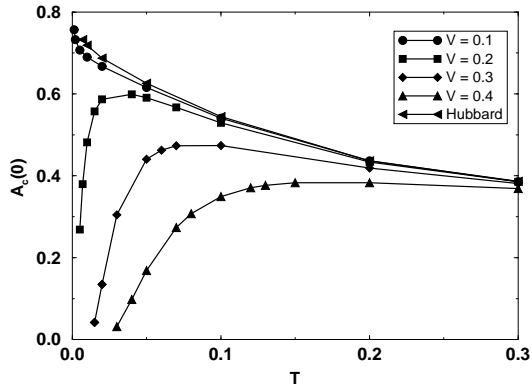,width=8cm,angle=270}}
\vskip 0.4cm
\caption{Spectral weight of the conduction electrons at the
chemical potential, $A_c(0)$, for $U_c = 1$ vs.\ temperature}
\label{qppeak_u1}
\end{figure}
The separation of the two subsystems is indicated in
Fig.~\ref{qppeak_u1} as well where we compare the spectral weight
at the chemical potential for different values of the $c$-$f$
hybridization $V$ and the pure $c$-Hubbard model ($V=0$): At high
to moderate temperatures the spectral weight of the
Anderson-Hubbard model follows the one of the pure Hubbard model.

In Fig.~\ref{qppeak_u1} we also see that this behavior does not
extend to low temperatures. At a certain temperature, which
depends on $V$, the spectral weight no longer follows the
quasiparticle peak of the Hubbard model but drops to zero. As
seen in Fig.~\ref{dos_u1v2} indeed a gap occurs and sharp
structures emerge close to the chemical potential when the
temperature falls below $T\lesssim 0.04$. This resembles the
Anderson model with uncorrelated conduction
electrons.\cite{Jarrell95} There, at $T<T_0$ where $T_0$ is a
characteristic temperature related to the Kondo temperature, the
Kondo effect leads to a resonance at the chemical
potential. These dynamically generated local states cross the
conduction-band states and one finds a splitting of the
conduction band with a gap at the position of the resonance. Due
to particle-hole symmetry, this feature occurs at the chemical
potential and the system becomes an insulator.\cite{Yamada85}

\begin{figure}
\centerline{\psfig{file=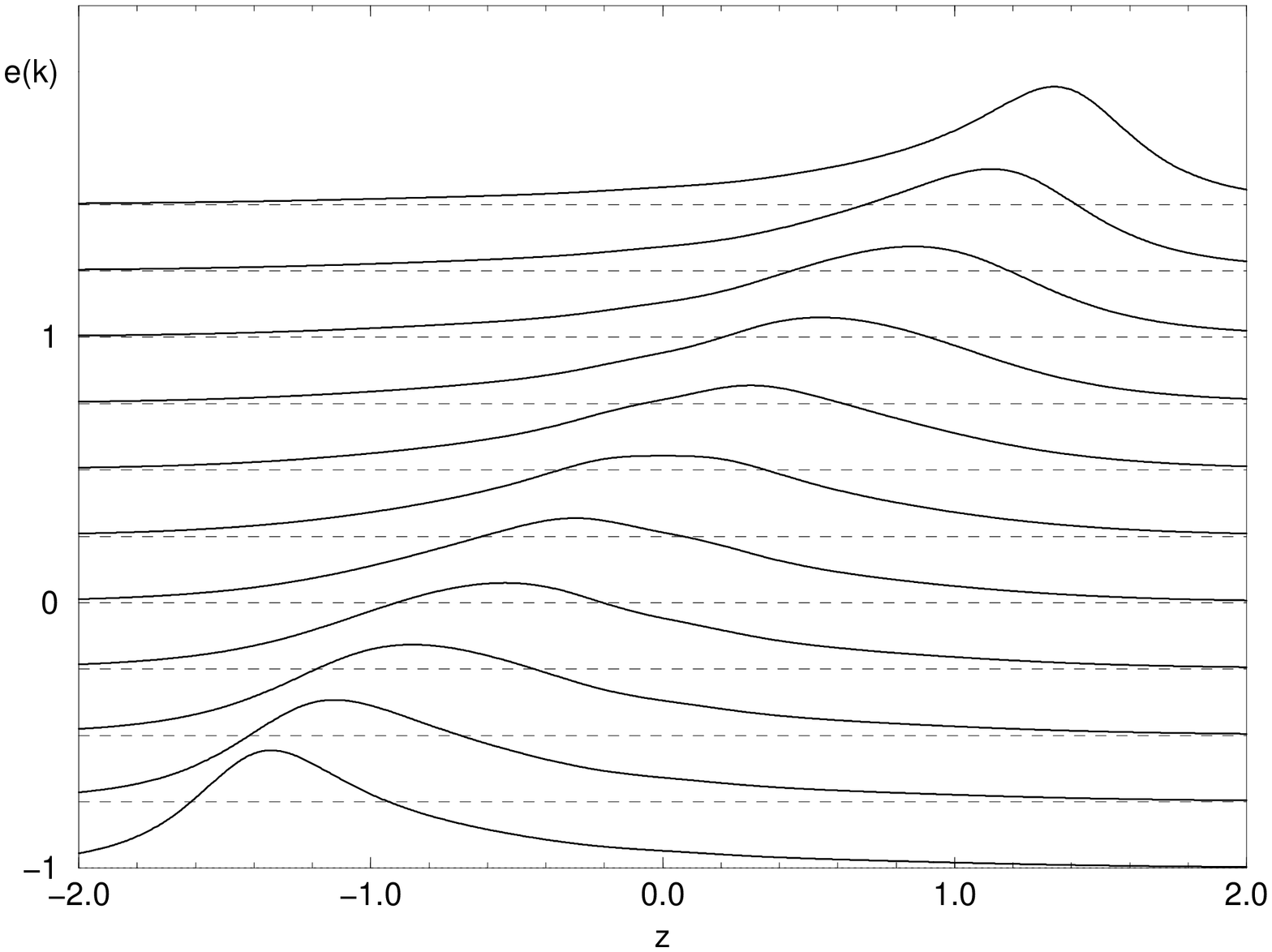,width=10cm,angle=270}}
\centerline{\psfig{file=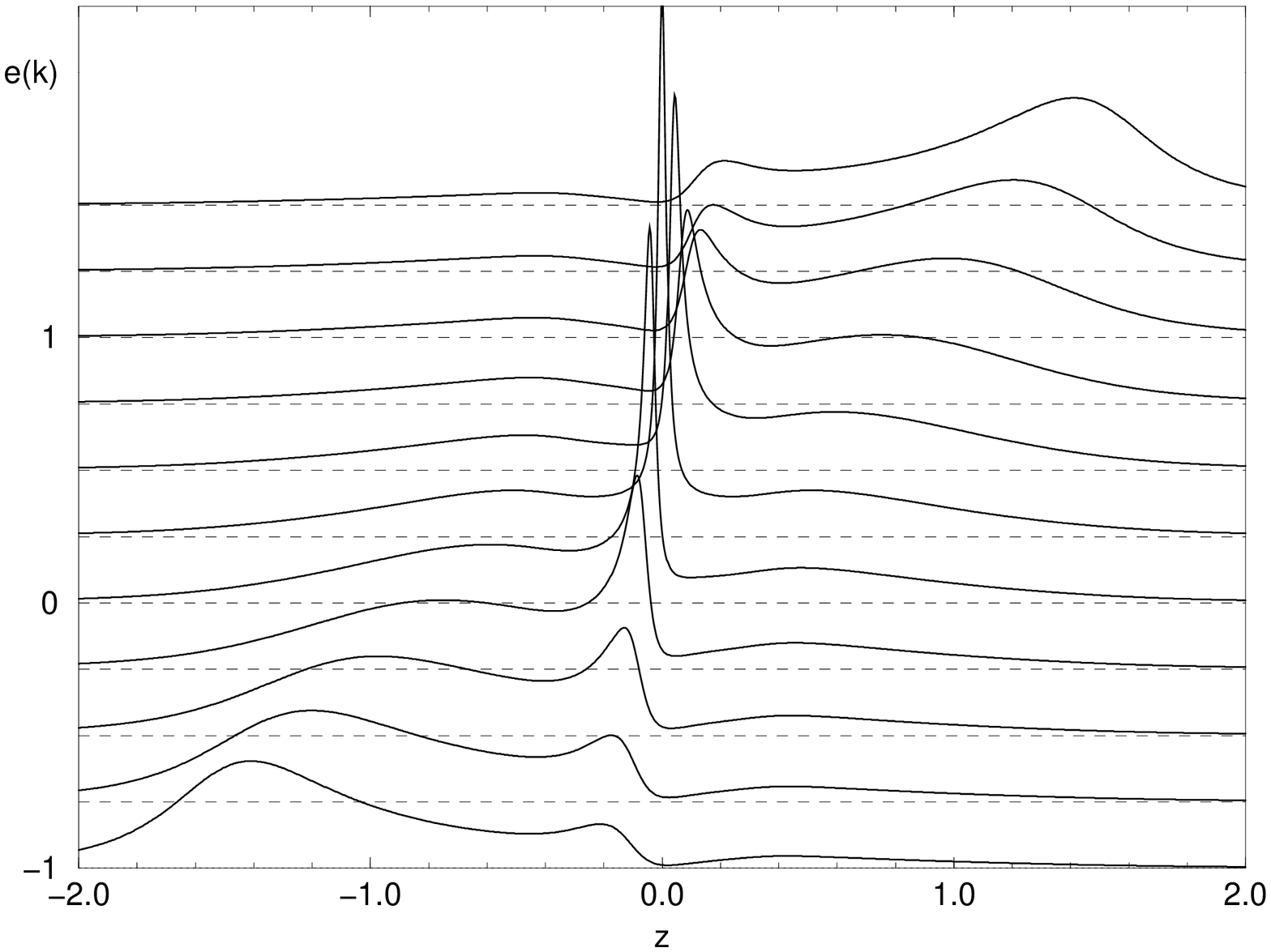,width=10cm,angle=270}}
\centerline{\psfig{file=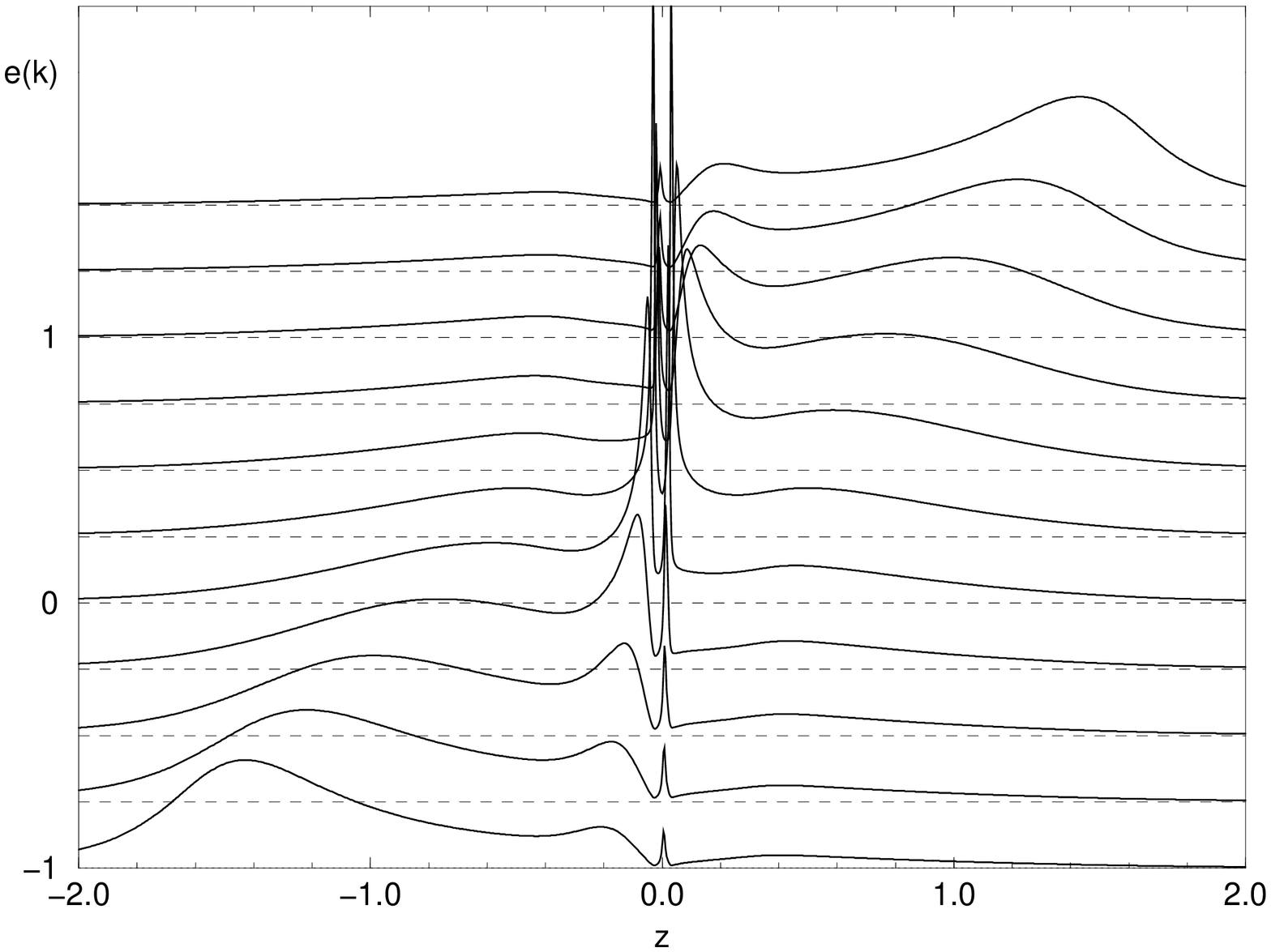,width=10cm,angle=270}}
\caption{Spectral function $A_c(k,z)$ for $U_c = 1$, $V = 0.2$
and different values of $\epsilon(k)$. {\em a.} $T=0.3$, {\em b.}
$T=0.05$, {\em c.} $T=0.005$} 
\label{spec_u1v2}
\end{figure}

A similar behavior is found in the case of interacting conduction
electrons. In order to see that indeed flat bands occur close to
the chemical potential we inspect the $k$-dependent spectral
function
\begin{equation}
A_c(k,z) = -\frac{1}{\pi} {\rm Im~} G_c(k,z) ~.
\end{equation}
From the self-consistency equation~(\ref{cond}) and
Eq.~(\ref{greens}) we find
\begin{equation}
G_c(k,z) = 
\frac{1}
{G_{{\rm loc},c}^{-1}(z) - \epsilon_k 
+ \frac{1}{4} G_{{\rm loc},c}(z)} ~. 
\end{equation} 
Since $A_c(k,z)$ depends on $k$ only via $\epsilon_k$, we plot
$A_c(k,z)$ for different $\epsilon_k = -1\dots 1$ in
Fig.~\ref{spec_u1v2}.
At high temperatures ($T = 0.3$, Fig.~\ref{spec_u1v2}{\em a.})
the features are very broad and we basically see the Hubbard
bands for the $c$-subsystem at $\sim U_c/2$ as well as for the
$f$-subsystem at $\sim U_f/2$ which weakly admix to the $c$
spectra. At intermediate temperatures ($T = 0.1\dots 0.02$,
$T=0.05$ is shown in Fig.~\ref{spec_u1v2}{\em b.}) spectral
weight is found at the chemical potential as well and we may
trace the quasiparticle band of the Hubbard model. When the
temperature is decreased below 0.015 ($T=0.005$,
Fig.~\ref{spec_u1v2}{\em c.}) these peaks split, a small gap
opens and the newly emerged peaks show a weak dispersion at the
chemical potential. The transition occurs rather quickly: Whereas
at $T=0.015$ only the peak at $\epsilon_k=0$ is split and the
band follows the quasiparticle band of the Hubbard band
elsewhere, two separated bands already emerged at $T=0.01$. Due
to their weak dispersion at the chemical potential we expect that
they will lead to heavy bands upon doping. At higher energies
these new bands merge in the previous quasiparticle bands of the
Hubbard model.

These findings fit qualitatively to the scenario of the Anderson
model with uncorrelated conduction electrons described
above\cite{Jarrell95} and to the results of the slaved-boson
mean-field treatment.\cite{Newns87} In the latter, the free
conduction bands hybridize with a strongly renormalized (reduced)
coupling to an effective $f$-level at the chemical
potential. This leads to weakly dispersive bands at low energies
merging in the original bands at high energies. In the
Anderson-Hubbard model the new bands merge in the former
quasiparticle band of the $c$-Hubbard model at high energies. We
conclude that these quasiparticles take the role of the free
conduction electrons in the dynamical screening of the
$f$-moments leading to the resonance at the chemical
potential. This interpretation does not contradict previous
results on a single impurity embedded in a correlated
host.\cite{Schork96} There it turned out that a variational
ansatz in the spirit of Varma and Yafet\cite{Varma76} which uses
quasiparticles for the screening of the $f$-moment instead of
bare electrons is not sufficient to find the correct Kondo
temperature. This holds due to importance of the renormalization
of the $c$-$f$ exchange interaction and does not imply that the
quasiparticle picture is not valid in describing the screening
process.

When the interaction strength of the conduction electrons is
increased to $U_c=1.5$, no qualitative changes occur at first
sight.
\begin{figure}
\centerline{\psfig{file=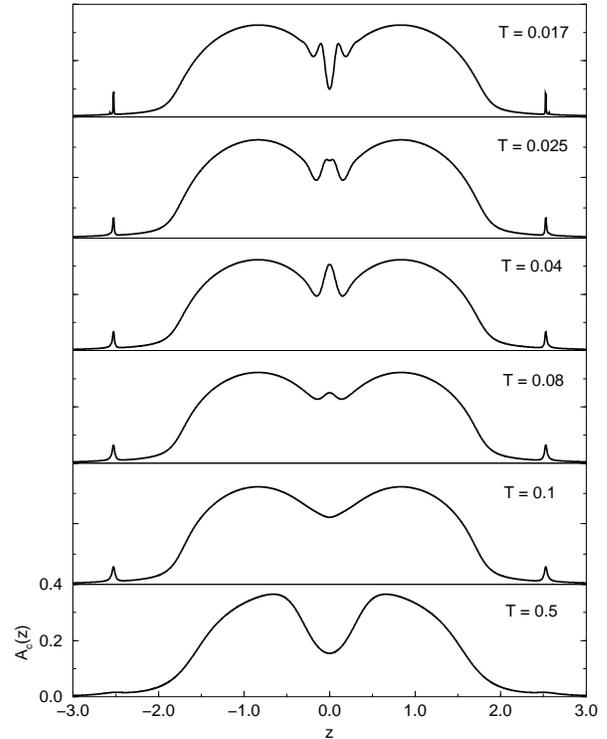,width=8cm}}
\caption{Spectral density $A_c(z)$ for $U_c = 1.5$, $V = 0.2$,
$U_f = 5$ at different temperatures}
\label{dos_u15v2}
\end{figure}
In Fig.~\ref{dos_u15v2} we show the $c$-spectra at various
temperatures for $U_c=1.5$ and $V=0.2$. Again, the $c$- and
$f$-subsystem are separated at high temperatures and a
quasiparticle peak of the $c$-Hubbard model evolves first. At low
temperatures a Kondo resonance is formed at the chemical
potential with hybridizes with the quasiparticle band, a gap
opens and we find bands with a weak dispersion. When compared to
$U_c=1$, we find that the gap has increased. This indicates that
the hybridization of the quasiparticle band with the dynamically
generated states has increased.

\begin{figure}
\centerline{\psfig{file=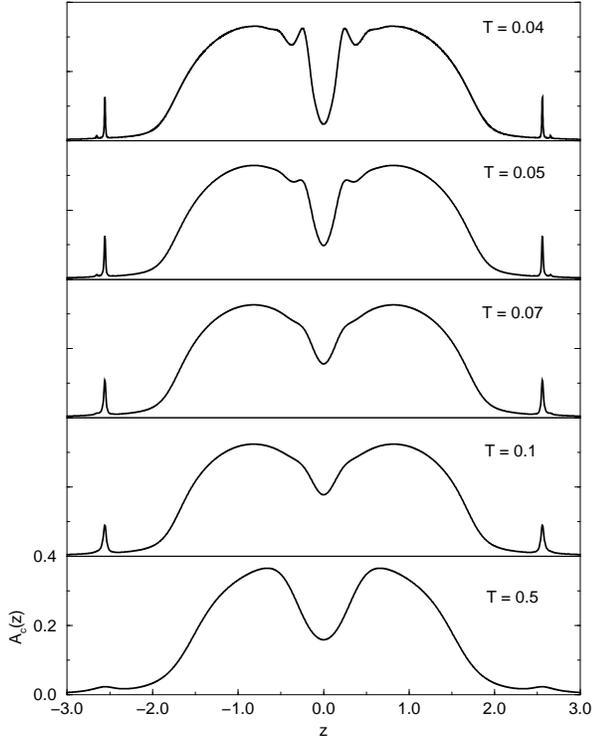,width=8cm}}
\caption{Spectral density $A_c(z)$ for $U_c = 1.5$, $V = 0.3$,
$U_f = 5$ at different temperatures}
\label{dos_u15v3}
\end{figure}
However, as demonstrated in Fig.~\ref{dos_u15v3}, a quasiparticle
peak does not always occur in the intermediate temperature range
even though the bare conduction electrons are metallic:
Increasing the $c$-$f$ mixing to $V=0.3$, the system becomes
directly insulating although $U_c = 1.5 < U_{\rm crit}$.
\begin{figure}
\centerline{\psfig{file=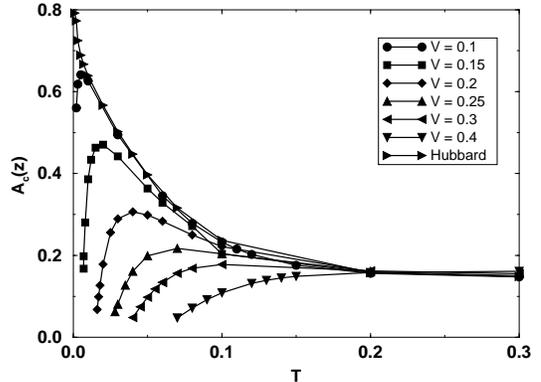,width=8cm,angle=270}}
\vskip 0.4cm
\caption{Spectral weight of the conduction electrons at the
chemical potential, $A_c(0)$, for $U_c = 1.5$ vs.\ temperature}
\label{qppeak_u15}
\end{figure}
Figure~\ref{qppeak_u15} illustrates that it depends on the value
of $V$ whether the quasiparticle peak shows up or not. We
conclude that part of the strong correlation on the $f$-orbital
is effectively inherited by the $c$-orbital via the hybridization
$V$. A similar effect has been observed and discussed for a
different model in Ref.~\onlinecite{Blawid96}.

Although no quasiparticle peak corresponding to the conduction
electrons emerges in Fig.~\ref{dos_u15v3}, we still recover the
Anderson scenario described above when decreasing the
temperature, i.e., a gap opens and peaks arise close to it at low
temperatures $T\sim 0.04$. It appears that we do not need
pre-formed quasiparticles in order to screen the $f$-moments
which in turn leads to the Kondo resonance. Note however, that
there is finite spectral weight at the chemical potential.
\begin{figure}
\centerline{\psfig{file=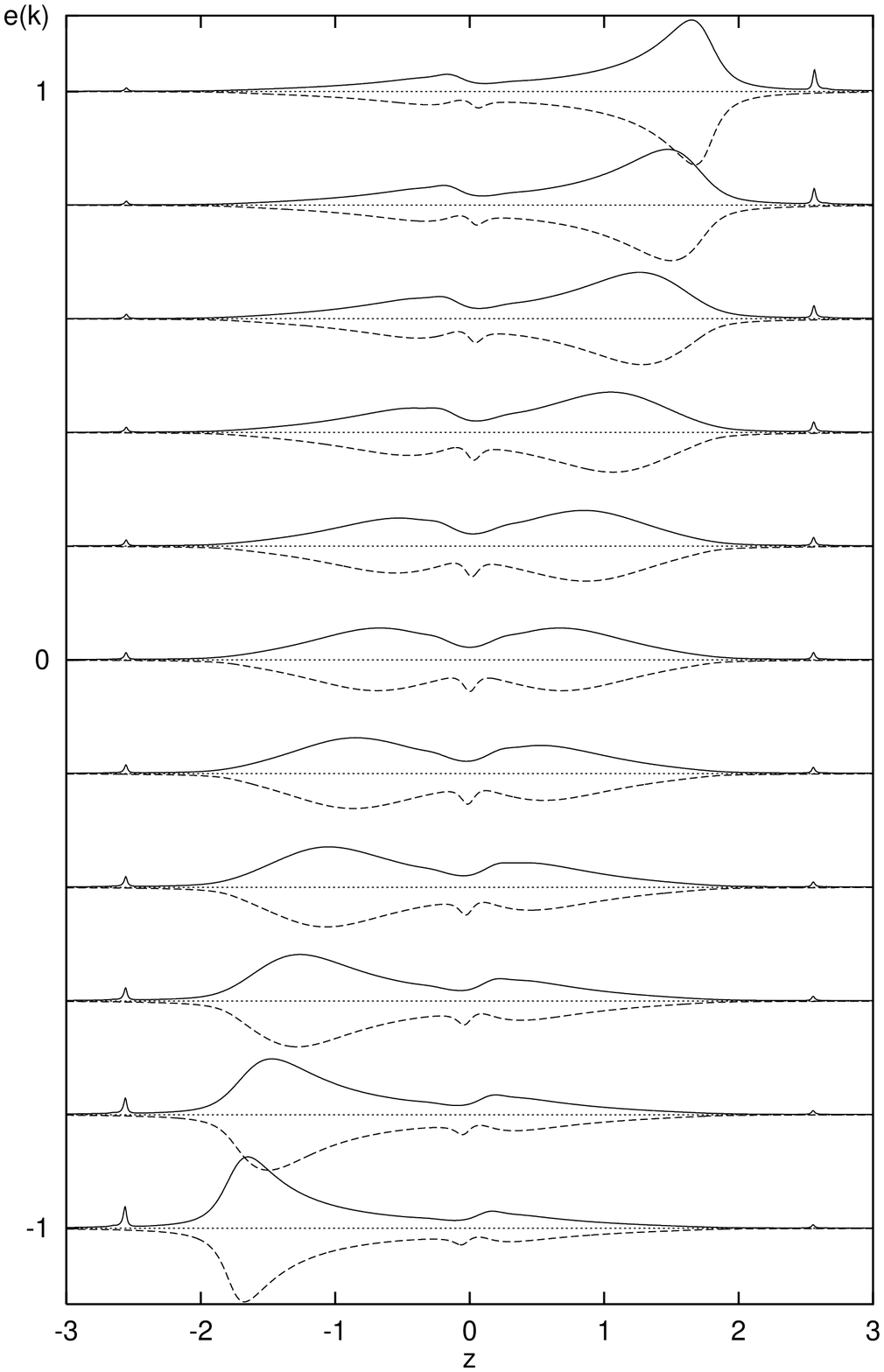,width=8cm,height=10.5cm}}
\centerline{\psfig{file=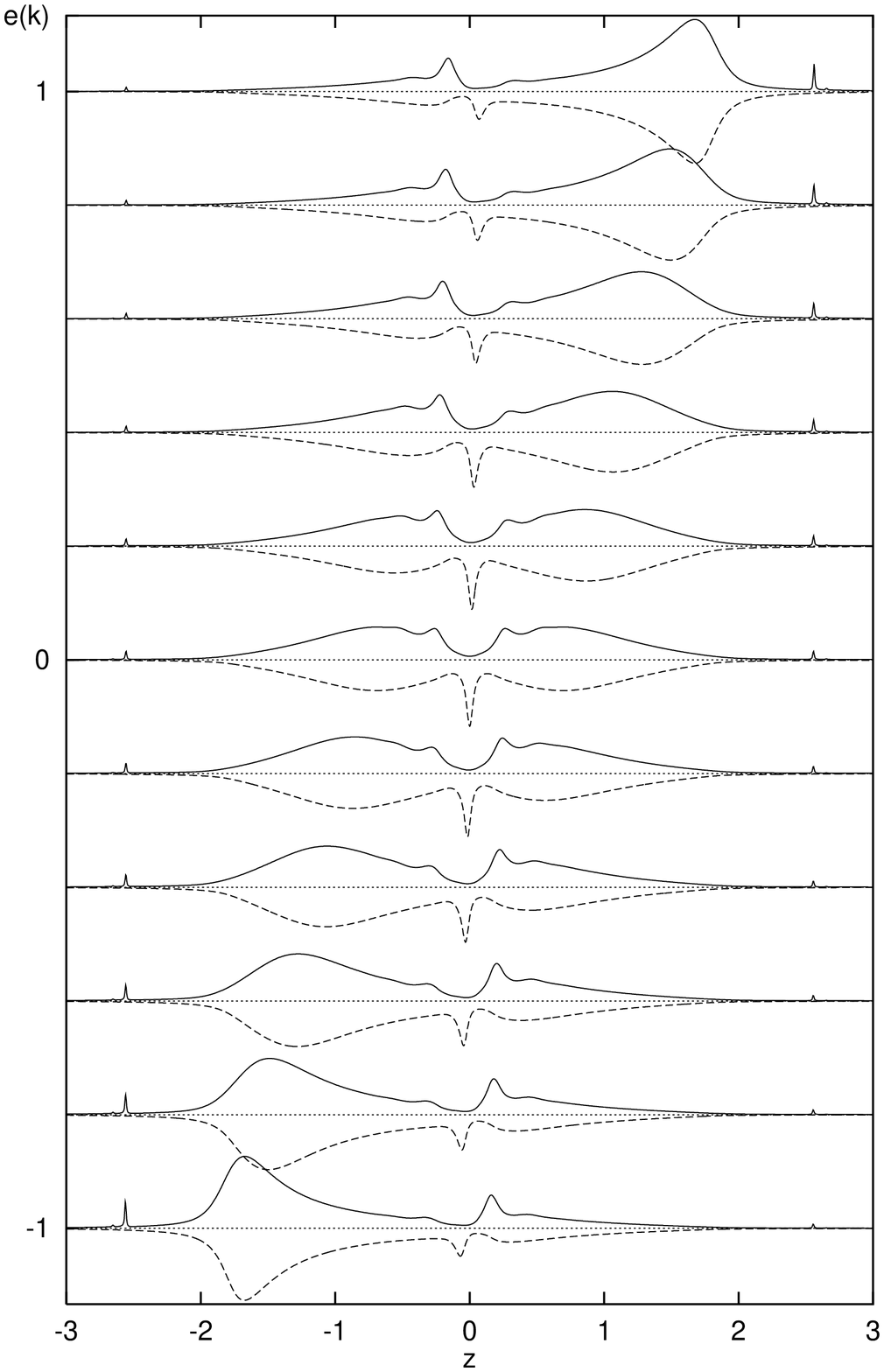,width=8cm,height=10.5cm}}
\caption{Comparison of the spectral function $A_c(k,z)$ of the
Anderson-Hubbard model ($U_c = 1.5$, $V=0.2$, solid lines) to the
Hubbard model ($U_c = 1.5$, dashed lines, multiplied by $-1$) at
{\em a.} $T=0.07$ and {\em b.} $T=0.04$.}
\label{spec_u15v3}
\end{figure}
This is better seen in the $k$-resolved spectral function
$A_c(k,z)$ in Fig.~\ref{spec_u15v3} where we display $A_c(k,z)$
for two different temperatures for the Anderson-Hubbard
($U_c=1.5, V=0.3$) and the Hubbard model ($U=1.5$). For the
Hubbard model we find a peak crossing the chemical potential at
$T=0.07$ (Fig.~\ref{spec_u15v3}{\em a.}). It corresponds to the
quasiparticle peak of the Hubbard model being formed at this
temperature. The Anderson-Hubbard model, however, exhibits no
structure crossing the chemical potential. Note that upper and
lower Hubbard bands of the Anderson-Hubbard model are in
agreement with those of the pure Hubbard model. At $T=0.04$
(Fig.~\ref{spec_u15v3}{\em b.})  the Anderson-Hubbard model shows
two flat bands at the chemical potential. The peaks are most
pronounced close to chemical potential. In contrast to the case
of small $U_c$ ($U_c=1$ discussed above) the corresponding bands
do not merge the quasiparticle band of the Hubbard model. This
indicates a large effective hybridization between quasiparticle
and dynamically generated states at the chemical potential. To a
lesser degree this behavior is also observed for $U_c=1.5$ and
$V=0.2$ ($A_c(k,z)$ not shown). We conclude that the effective
hybridization increases as both $U_c$ and $V$
increase. Considering again Fig.~\ref{spec_u15v3}{\em a.}, we
find that the difference between Hubbard and Anderson-Hubbard
model at high temperatures is related to the onset of the heavy
bands which are thus already seen at comparatively high
temperatures.

We finally turn to the case where the bare $c$-subsystem is
insulating, i.e., $U_c>U_{\rm crit}$.
\begin{figure}
\centerline{\psfig{file=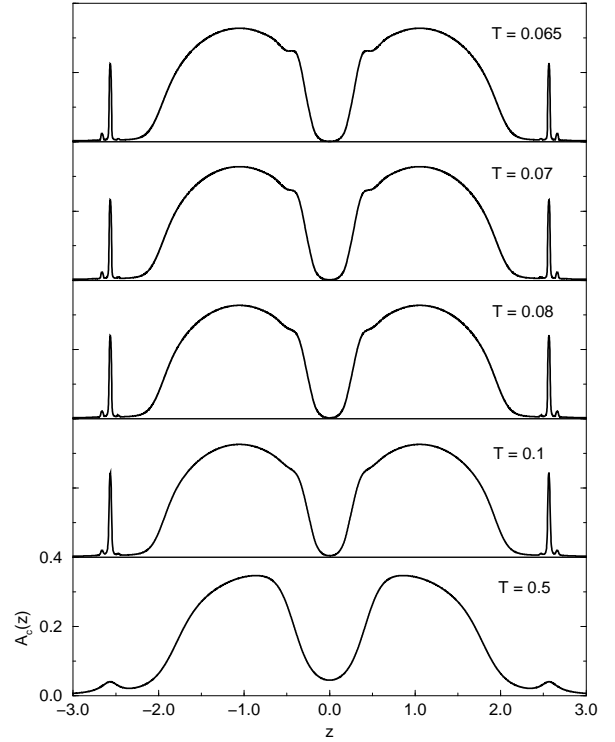,width=8cm}}
\caption{Spectral density $A_c(z)$ for $U_c = 2$, $V = 0.3$, 
$U_f = 5$} 
\label{dos_u2v3}
\end{figure}
For $U_c =2$ a shoulder in the spectral function emerges at the
edge Hubbard band towards the chemical potential when the
temperature is lowered (Fig.~\ref{dos_u2v3}).
\begin{figure}
\centerline{\psfig{file=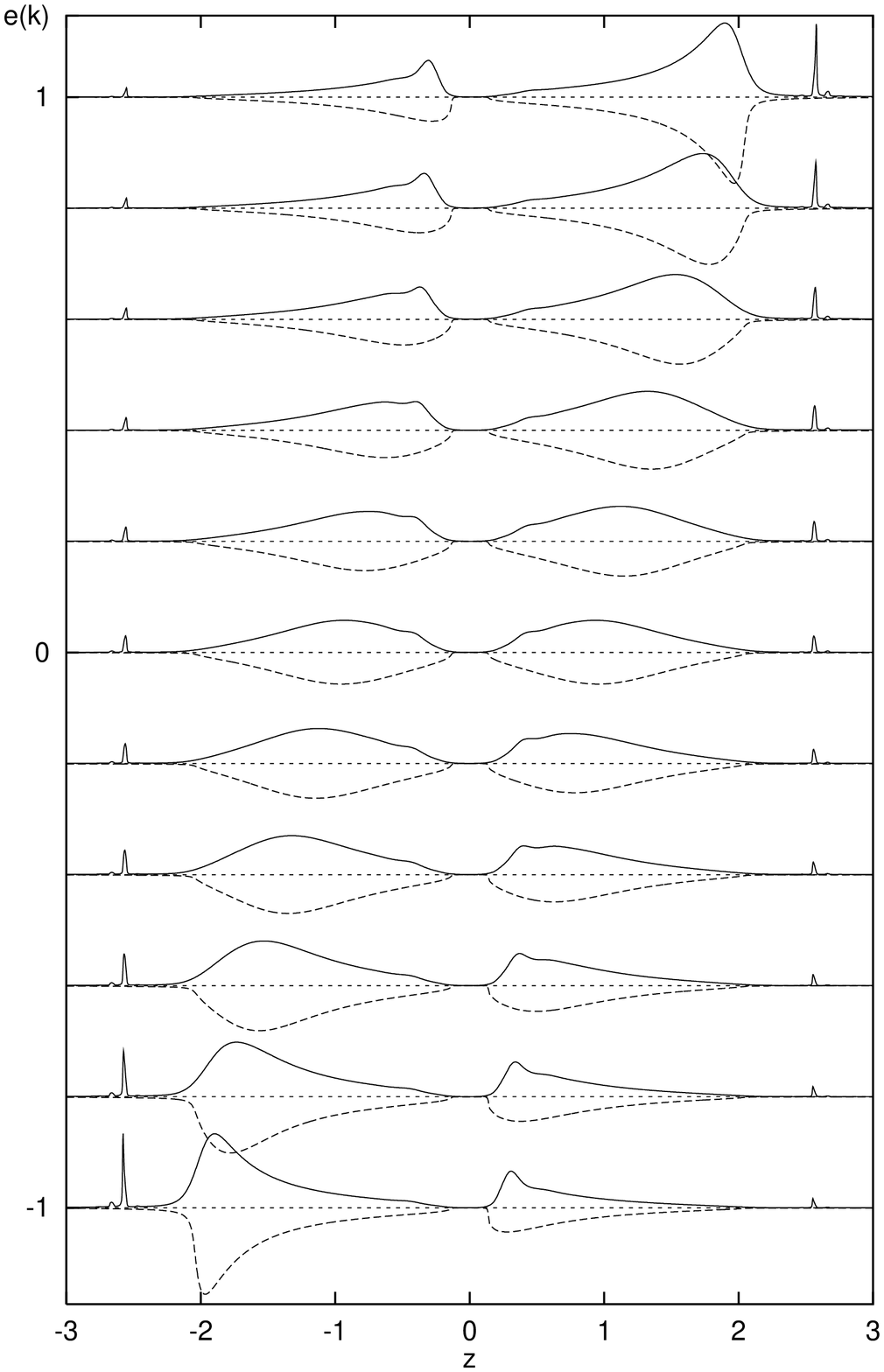,width=8cm}}
\caption{Comparison of the $c$ spectral function $A_c(k,z)$ of
the Anderson-Hubbard model ($U_c = 2$, $V = 0.3$, $U_f = 5$,
solid lines) to the Hubbard model ($U_c=2$, dashed lines,
multiplied by $-1$) at $T=0.065$}
\label{spec_u2v3}
\end{figure}
That this feature is indeed caused by the $f$-subsystem is
demonstrated in Fig.~\ref{spec_u2v3} where we compare $A_c(k,z)$
for the Anderson-Hubbard model and the pure Hubbard
model. Whereas the structures corresponding to the Hubbard bands
roughly agree, the Anderson-Hubbard model shows an additional
peak at the low-energy edge of the Hubbard bands. This is
unexpected because the $c$-subsystem is insulating and provides
no quasiparticles which could screen the $f$-moment and the
resulting spectra should resemble the one of the $c$-$f$ molecule
described in Sec.~\ref{subsec:local}.  We do not believe that
this shoulder is spurious since increasing the energy resolution
which allows to proceed to lower temperatures did not change the
shoulder. However, we can not exclude a principle failure of the
extended NCA as it is well-known that the NCA fails to converge
when the system becomes insulating.\cite{Pruschke93} On the other
hand, the spectral weight at the chemical potential is not
exactly zero. Similar to the case $U_c=1.5$, $V=0.3$ discussed
above, this small, but finite spectral weight could be sufficient
to dynamically generate states at the chemical potential which
result in a heavy band via hybridization as previously. To
clarify this situation, it is mandatory to employ other numerical
methods for solving the impurity model. If the picture presented
above is valid, the effective hybridization must be large
compared to the metallic cases so that the new bands are pushed
towards the Hubbard bands. Note that the new peaks approach, but
not merge in the Hubbard bands whereas they merged in the
quasiparticle bands in the metallic cases ($U_c=1$, and
$U_c=1.5$, $V\lesssim 0.2$).

%%%%%%%%%%%%%%%%%%%%%%%%%%%%%%%%%%%%%%%%%%%%%%%%%%%%%%%%%%%%%%%%%%
%%%%%%   HYBRIDIZATION GAP   %%%%%%%%%%%%%%%%%%%%%%%%%%%%%%%%%%%%%
%%%%%%%%%%%%%%%%%%%%%%%%%%%%%%%%%%%%%%%%%%%%%%%%%%%%%%%%%%%%%%%%%%

\subsection{Hybridization gap}
\label{subsec:gap}

\begin{figure}
\centerline{\psfig{file=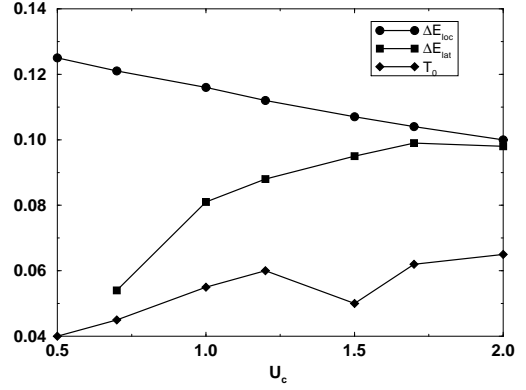,width=8cm,angle=270}}
\vskip 0.4cm
\caption{Splitting of the $f$ peak in the molecule and lattice,
and $T_0$ vs.\ $U_c$ at $V=0.3$} 
\label{delta_e}
\end{figure}

In this section we focus on the gap which opens at low
temperatures. We determine the size of the gap, $\Delta$, as
twice the distance between zero frequency and the first maximum
close to the chemical potential.\cite{Jarrell95}
\begin{figure}
\centerline{\psfig{file=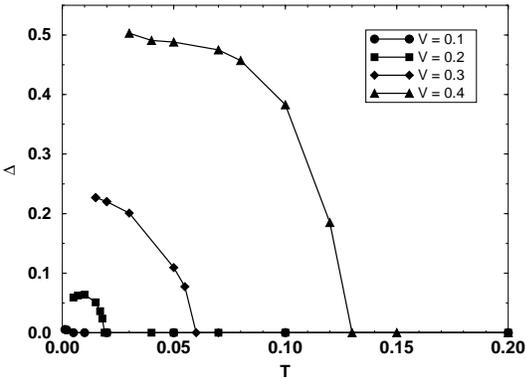,width=8cm,angle=270}}
\vskip 0.4cm
\caption{Gap in the spectral density vs.\ temperature. The gaps
are obtained from the two maxima in $A_c(z)$ for $U_c = 1$ and
different hybridizations $V$}
\label{gap_u1}
\end{figure}
In Fig.~\ref{gap_u1}, where we plot the gap $\Delta$ vs.\
temperature for different hybridizations $V$, we observe that the
gap opens quickly when the temperature is reduced and we define
$T_0$ as the temperature where the gap opens. One expects that
$T_0$ is related to the energy difference, $\Delta E$, of singlet
and triplet states in the local problem, see
Sec.~\ref{subsec:local}. Indeed, we observe in
Figs.~\ref{dos_u15v2} and~\ref{dos_u15v3} that the gap opens
roughly in the same temperature regime where the $f$-peak at $z
\sim \pm U_f/2$ splits, i.e., where the states $|\Psi_{\rm
S}\rangle$ and $|\Psi_{\rm T}\rangle$ become distinguishable in
the photoemission of the lattice. We extract this splitting of
the $f$-peak, $\Delta E_{\rm lat}$ from the spectral function of
the lattice model (if visible) and compare it to the
corresponding splitting, $\Delta E_{\rm loc}$ for the $c-f$
molecule [cf.\ Eq.~(\ref{exc_energy})] in Fig.~\ref{delta_e} for
different values of $U_c$ and $V=0.3$.
Surprisingly we find that while both splittings are of the same
order of magnitude, they depend differently on $U_c$: When $U_c$
increases, $\Delta E_{\rm lat}$ increases, whereas $\Delta E_{\rm
loc}$ decreases.

\begin{figure}
\centerline{\psfig{file=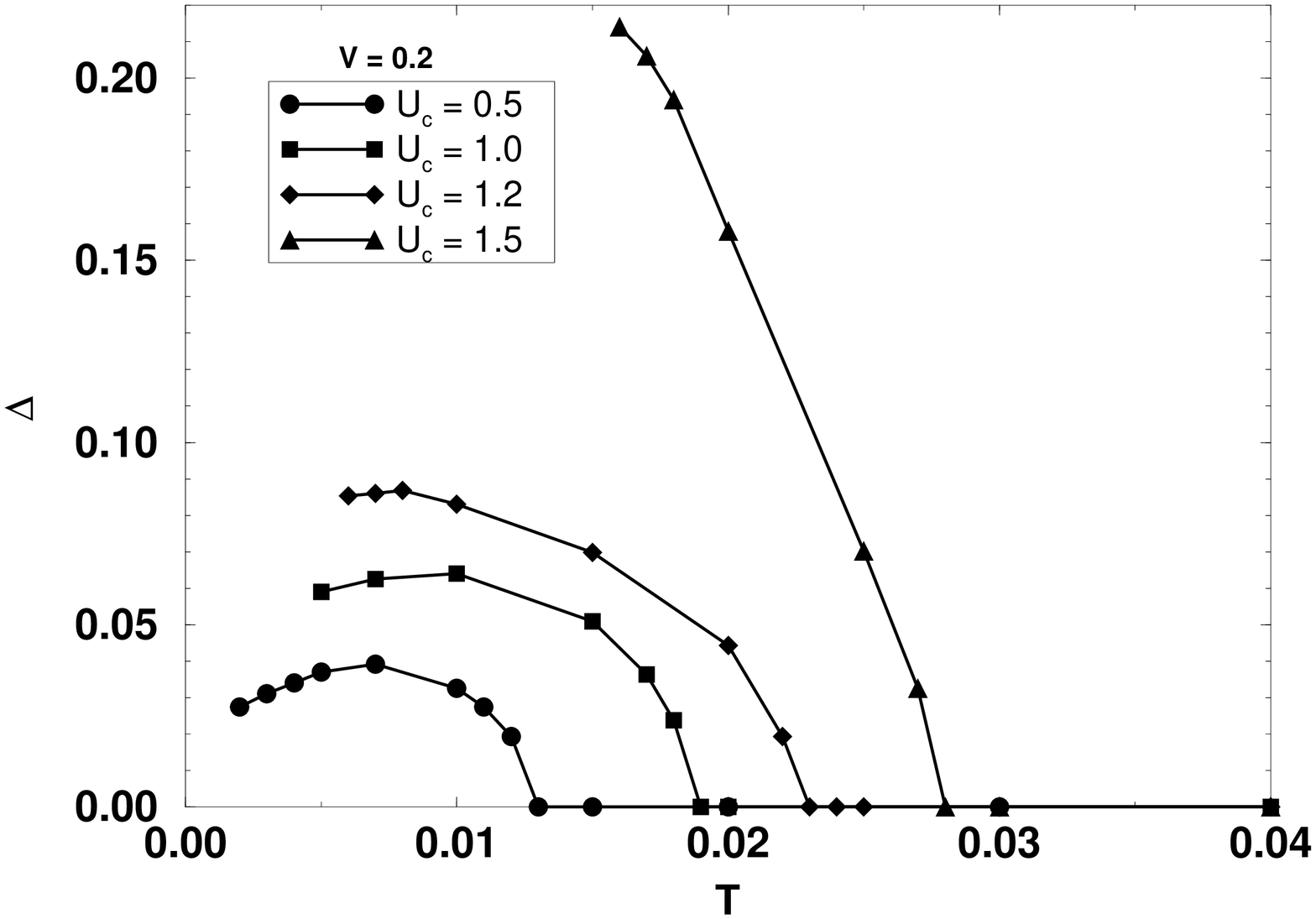,width=8cm,angle=270}}
\centerline{\psfig{file=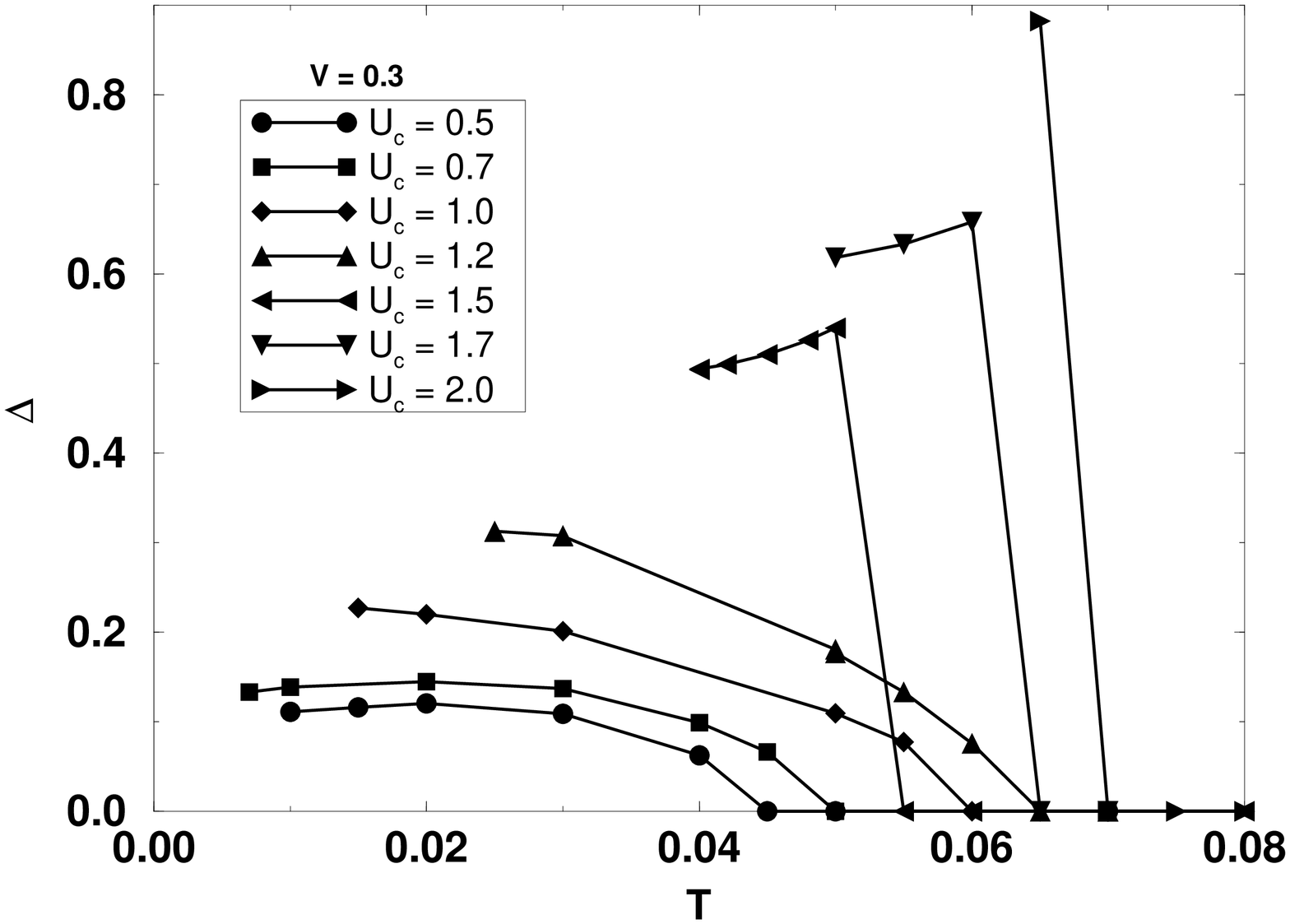,width=8cm,angle=270}}
\vskip 0.4cm
\caption{Gap in the spectral density vs.\ temperature. The gaps
are obtained from the two maxima in $A_c(z)$ for different
interaction strength $U_c$ and {\em a.} $V = 0.2$ and {\em b.}
$V=0.3$}
\label{gap_v3}
\end{figure}
In Fig.~\ref{gap_v3} we plot the gap vs.\ temperature for
different values of $U_c$ and $V = 0.2,~0.3$. In general, the
temperature $T_0$ at which the gap sets in increases with $U_c$,
but deviations occur: When $V=0.3$, $U_c = 1.5$ and $1.7$ do not
fit into this scheme. However, one should bear in mind that the
onset of the gap formation as we measure it, depends also on the
shape of the spectrum. As is seen from Fig.~\ref{delta_e}, $T_0$
is of the same order of magnitude as $\Delta E_{\rm lat}$ and its
general behavior agrees with $\Delta E_{\rm lat}$ and is thus
opposed to $\Delta E_{\rm loc}$.

We now turn to the size of the gap. It has been shown for the
Anderson model with free conduction electrons that this quantity
determines the low-temperature thermodynamics in the limit of
infinite dimensions.\cite{Mutou94,Mutou95,Jarrell95} The gaps in
the local spin and charge excitation spectra were found to be of
the same order of magnitude as the gap in the density of
states. The latter thus provides a measure of the low-temperature
scale (``Kondo temperature''). From Fig.~\ref{gap_v3} we extract
that the size of the gap has not yet converged at the lowest
temperatures where our extended NCA ceases to
converge. Nevertheless, we can draw some qualitative conclusions:
The size of the gap increases systematically as $U_c$ increases.
\begin{figure}
\centerline{\psfig{file=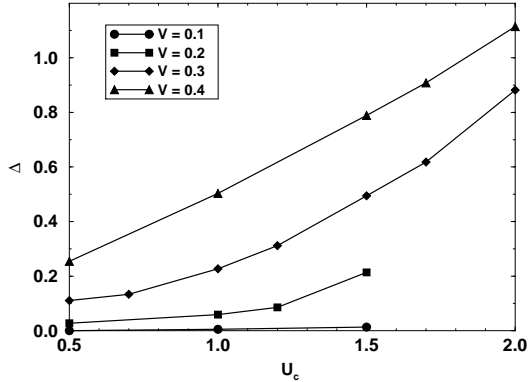,width=8cm,angle=270}}
\vskip 0.4cm
\caption{Gap in the spectral density vs.\ $U_c$. The gaps
correspond to the lowest temperatures reached.}
\label{gap_vs_U}
\end{figure}
This is also deduced from Fig.~\ref{gap_vs_U} where we plotted
the gap at the lowest temperature we could reach for each pair
$(U_c,V)$ vs.\ the strength $U_c$ of the Hubbard interaction of
the conduction electrons. This procedure implies that the points
shown correspond to different temperatures. Note that the
magnitude of $\Delta$ varies much stronger with $U_c$ and $V$
compared to $T_0$ or $\Delta E_{\rm }$.

%%%%%%%%%%%%%%%%%%%%%%%%%%%%%%%%%%%%%%%%%%%%%%%%%%%%%%%%%%%%%%%%%
%%%%%%   CONCLUSIONS   %%%%%%%%%%%%%%%%%%%%%%%%%%%%%%%%%%%%%%%%%%
%%%%%%%%%%%%%%%%%%%%%%%%%%%%%%%%%%%%%%%%%%%%%%%%%%%%%%%%%%%%%%%%%

\section{Conclusions}
\label{sec:conclusions}

In conclusion, we studied the influence of interactions among the
conduction electrons on the low-temperature behavior of the
periodic Anderson model. In the dynamical mean-field theory the
model is mapped onto a generalized Anderson impurity model that
couples the orbitals of a single unit cell to an effective medium
which has to be determined self consistently. The impurity model
was solved numerically by an extended non-crossing approximation.

For weakly interacting conduction electrons we found that at high
temperatures the $c$- and $f$-subsystems are almost separated as
in the case of free conduction electrons. Decreasing temperature
then first leads to the formation of quasiparticles in the
$c$-subsystem as in the bare Hubbard model. When the temperature
is reduced further, the quasiparticle band splits, a tiny gap
opens and the system turns into an insulator. As in the case of
free conduction electrons, the gap is formed by the level
crossing of the quasiparticle (Hubbard) band and the resonance
which arises at the chemical potential from the Kondo-like
screening of the $f$-moments. We observed that the quasiparticles
play an essential role in this screening. The resulting two bands
have a weak dispersion close to the gap and we expect that they
turn into heavy bands upon doping.

When the correlations of the conduction electrons become
stronger, the low-temperature gap increases. This can be
interpreted as increasing the effective hybridization between the
quasiparticle states at the chemical potential and the
dynamically generated states which leads to a larger gap and is
qualitatively in agreement with results found for impurity
models. For the latter case, the main effect of the (small)
interaction was to renormalize and increase the exchange
interaction.\cite{Schork96}

It turned out, however, that pre-formed quasiparticles within the
$c$-subsystem are not prerequisite for the emergence of heavy
bands. When increasing the $c$-$f$ hybridization, the $c$
orbitals seem to inherit correlations from the $f$ orbitals and a
quasiparticle peak is no longer formed in the spectral density at
intermediate temperatures, although the spectral weight at the
chemical potential does not vanish. Nevertheless, we observed
heavy bands at low temperatures in the one-particle spectra.

Even when the (bare) $c$-system is insulating, it is influenced
by the $f$-system at low temperatures: A shoulder forms at the
edge of the Hubbard band which shows weak dispersion. This is
surprising since the bare $c$-system provides no quasiparticles
which could screen the $f$-moments and the $c$ spectral weight
close to the chemical potential is small. However, we can not
exclude that this result is an artefact of our method to solve
the impurity problem and further investigations are necessary to
decide upon this question.

Finally, we investigated the hybridization gap which occurs at
low temperatures. When increasing the Coulomb interaction among
the conduction electrons the temperature at which the gap opens
increases. This temperature is related to the splitting of the
$f$-peak in the spectrum which results from singlet and triplet
states in the $c$-$f$ molecule. However, we found that this
splitting depends oppositely on the interaction strength in the
lattice and in the molecule. The low-temperature thermodynamics
in infinite dimensions scales with the size of the
gap.\cite{Jarrell95} This quantity therefore provides a measure
for the ``Kondo temperature.'' In agreement with impurity models,
this gap increases as the correlations among the conduction
electrons become stronger.

\acknowledgements

We would like to acknowledge useful discussions with K.\ Fischer,
P.\ Fulde, J.\ Keller, and J.\ Schmalian.

%%%%%%%%%%%%%%%%%%%%%%%%%%%%%%%%%%%%%%%%%%%%%%%%%%%%%%%%%%%%%%%%%
%%%%%%   REFERENCES   %%%%%%%%%%%%%%%%%%%%%%%%%%%%%%%%%%%%%%%%%%%
%%%%%%%%%%%%%%%%%%%%%%%%%%%%%%%%%%%%%%%%%%%%%%%%%%%%%%%%%%%%%%%%%


\begin{thebibliography}{10}

\bibitem{Fulde88}
P. Fulde, J. Keller, and G. Zwicknagl,  in {\em Solid state physics}, edited by
  H. Ehrenreich and D. Turnbell (Academic Press, San Diego, 1988), Vol.~41,
  pp.\ 1--150.

\bibitem{HewsonBook}
A.~C. Hewson, {\em The Kondo Problem to Heavy Fermions} (Cambridge University
  Press, Cambridge, 1993).

\bibitem{Brugger93}
T. Brugger {\it et~al.}, Phys. Rev. Lett. {\bf 71},  2481  (1993).

\bibitem{Fulde93}
P. Fulde, V. Zevin, and G. Zwicknagl, Z. Phys. B {\bf 92},  133  (1993).

\bibitem{Skanthakumar89}
S. Skanthakumar {\it et~al.}, Physica C {\bf 160},  124  (1989).

\bibitem{Oseroff90}
S.~B. Oseroff {\it et~al.}, Phys. Rev. B {\bf 41},  1934  (1990).

\bibitem{Furusaki94}
A. Furusaki and N. Nagaosa, Phys. Rev. Lett. {\bf 72},  892  (1994).

\bibitem{Li95}
Y.~M. Li, Phys. Rev. B {\bf 52},  R6979  (1995).

\bibitem{Froejdh96}
P. Fr{\"o}jdh and H. Johannesson, Phys. Rev. B {\bf 53},  3211  (1996).

\bibitem{Schork94}
T. Schork and P. Fulde, Phys. Rev. B {\bf 50},  1345  (1994).

\bibitem{Khaliullin95}
G. Khaliullin and P. Fulde, Phys. Rev. B {\bf 52},  9514  (1995).

\bibitem{Igarashi95}
J. Igarashi, T. Tonegawa, M. Kaburagi, and P. Fulde, Phys. Rev. B {\bf 51},
  5814  (1995).

\bibitem{Igarashi95b}
J. Igarashi, K. Murayama, and P. Fulde, Phys. Rev. B {\bf 52},  15966  (1995).

\bibitem{Schork96}
T. Schork, Phys. Rev. B {\bf 53},  5626  (1996).

\bibitem{Wang96b}
X. Wang, preprint (unpublished).

\bibitem{Shibata96}
N. Shibata, T. Nishino, K. Ueda, and C. Ishii, Phys. Rev. B {\bf 53},  R8828
  (1996).

\bibitem{Itai96}
K. Itai and P. Fazekas, Phys. Rev. B {\bf 54},  R752  (1996).

\bibitem{Metzner89}
W. Metzner and D. Vollhardt, Phys. Rev. Lett. {\bf 62},  324  (1989).

\bibitem{MuellerHartmann89}
E. M{\"u}ller-Hartmann, Z. Phys. B {\bf 74},  507  (1989).

\bibitem{Georges96}
A. Georges, G. Kotliar, W. Krauth, and M. Rozenberg, Rev. Mod. Phys. {\bf 68},
  13  (1996).

\bibitem{Ohkawa91}
F.~J. Ohkawa, J. Phys. Soc. Jpn. {\bf 60},  3218  (1991).

\bibitem{Jarrell92}
M. Jarrell, Phys. Rev. Lett. {\bf 69},  168  (1992).

\bibitem{Georges92b}
A. Georges, G. Kotliar, and Q. Si, Int. J. Mod. Phys. {\bf 6},  705  (1992).

\bibitem{Ohkawa92}
F.~J. Ohkawa, J. Phys. Soc. Jpn. {\bf 61},  1615  (1992).

\bibitem{Georges92}
A. Georges and G. Kotliar, Phys. Rev. B {\bf 45},  6479  (1992).

\bibitem{Ohkawa92c}
F.~J. Ohkawa, Phys. Rev. B {\bf 46},  9016  (1992).

\bibitem{Jarrell93b}
M. Jarrell, H. Akhlaghpour, and T. Pruschke, Phys. Rev. Lett. {\bf 70},  1670
  (1993).

\bibitem{Jarrell95}
M. Jarrell, Phys. Rev. B {\bf 51},  7429  (1995).

\bibitem{Saso96}
T. Saso and M. Itoh, Phys. Rev. B {\bf 53},  6877  (1996).

\bibitem{Keiter70}
H. Keiter and J.~C. Kimball, Phys. Rev. Lett. {\bf 25},  672  (1970).

\bibitem{Bickers87}
N.~E. Bickers, Rev. Mod. Phys. {\bf 59},  845  (1987).

\bibitem{Pruschke89}
T. Pruschke and N. Grewe, Z. Phys. B {\bf 74},  439  (1989).

\bibitem{Lombardo96}
P. Lombardo, M. Avignon, J. Schmalian, and K.-H. Bennemann, Phys. Rev. B {\bf
  54},  5317  (1996).

\bibitem{AbrikosovBook}
A.~A. Abrikosov, L.~P. Gorkov, and I.~E. Dzialoshinski, {\em Methods of Quantum
  Field Theory in Statistical Physics} (Pergamon, Elmsford, NY, 1965).

\bibitem{Si93}
Q. Si and G. Kotliar, Phys. Rev. B {\bf 48},  13881  (1993).

\bibitem{Bickers87b}
N.~E. Bickers, D.~L. Cox, and J.~W. Wilkins, Phys. Rev. B {\bf 36},  2036
  (1987).

\bibitem{Tsvelick83}
A.~M. Tsvelick and P.~B. Wiegmann, Adv. Phys. {\bf 32},  453  (1983).

\bibitem{Rice86}
T.~M. Rice and K. Ueda, Phys. Rev. B {\bf 34},  6420  (1986).

\bibitem{Yamada85}
K. Yamada and K. Yoshida,  in {\em Theory of Heavy Fermions and Valence
  Fluctuations}, Vol.~62 of {\em Springer Series in Solid-State Sciences},
  edited by T. Kasuya and T. Saso (Springer, Berlin, 1985), pp.\ 183--194.

\bibitem{Newns87}
D.~M. Newns and N. Read, Adv. Phys. {\bf 36},  799  (1987).

\bibitem{Varma76}
C.~M. Varma and Y. Yafet, Phys. Rev. B {\bf 13},  2950  (1976).

\bibitem{Blawid96}
S. Blawid, H.~A. Tuan, T. Yanagisawa, and P. Fulde, Phys. Rev. B {\bf 54},
  7771  (1996).

\bibitem{Pruschke93}
T. Pruschke, D.~L. Cox, and M. Jarrell, Phys. Rev. B {\bf 47},  3553  (1993).

\bibitem{Mutou94}
T. Mutou and D.~S. Hirashima, J. Phys. Soc. Jpn. {\bf 63},  4475  (1994).

\bibitem{Mutou95}
T. Mutou and D.~S. Hirashima, J. Phys. Soc. Jpn. {\bf 64},  4799  (1995).

\end{thebibliography}
\end{document}